\documentclass[12pt,twoside]{article}
\usepackage{fleqn,espcrc1}
\usepackage{graphicx,epsfig}

\newcommand{\AmS}{{\protect\the\textfont2
  A\kern-.1667em\lower.5ex\hbox{M}\kern-.125emS}}

\title{Strangeness Production by Electromagnetic and Hadronic Probes
       \thanks{Work supported by French Atomic Energy Commission and partly
        by European Commission under Contract HPRN-CT-2000-00130.}
}

\author{J.-M. Laget \\
        CEA/SACLAY, DAPNIA/SPhN,   
        F91191 Gif-Sur-Yvette, France}
 
\begin{document}

\maketitle

\begin{abstract}
After pioneering works on hypernuclei, strangeness production mechanisms have been studied in hadron collisions and photoreactions in the sixties. Recent experiments at SATURNE and COSY,
in the hadronic sector, as well as ELSA and JLab, in the electromagnetic sector, have confirmed our basic ideas on  the reaction mechanisms. In the near future, strangeness production at JLab,
HERMES and COMPASS may prove to be a powerful tool to study hadronic matter. 
\end{abstract}

\section{Introduction}

Since the strange quark is not a normal building block of nucleons and nuclei, strangeness production is a powerful tool to study properties of Hadronic Matter. As in the study of any complex system, the inclusion of an ``impurity'' and the study of its subsequent propagation provides us with a way to reveal configurations or states that can not be reached otherwise. In the fifties, pioneering studies of hypernuclei led to a first estimate of the Nucleon-Hyperon scattering lengths, while more recent studies revealed the inner shell structure of heavy hypernuclei (see Ref~\cite{Be91} for a review). The world data set on strangeness production in $pp$ scattering was very meager~\cite{Fla79}, until SATURNE was able to accurately determine cross sections and spin observables. Now COSY is in the process of producing a comprehensive set of data near and just above threshold.

In the electromagnetic sector, the sparse data which have been collected in the fifties and  early sixties are going to be superseded by works ongoing at ELSA and JLab.

In the written version of this talk, I will concentrate on new (theoretical and experimental) developments, referring the reader to a lecture~\cite{La93} and a recent paper~\cite{La97} for a general background and reference to older experiments. I will start with the Hadronic Sector, deal with the Electromagnetic Sector, discuss what can we learn in the Few Body Systems and end my talk discussing issues in hidden strangeness production. Hypernuclear physics is covered elsewhere in these proceedings.

\section{Hadronic Sector}

The world set of total cross-sections of the three channels $pp \rightarrow p K^+ Y$, ($Y= \Lambda, \Sigma^0, \Sigma^+$), is very well reproduced over the energy range $2\leq T_P \leq 6$ GeV by the simplest model~\cite{La91} which assumes $\pi$ and $K$ meson exchange, supplemented by Nucleon Hyperon rescattering (FSI). The model relies on the experimental amplitudes of each elementary subprocess (elastic Kaon scattering, Hyperon production by pion beams), Hyperon Nucleon coupling constants consistent with SU3, and the Hyperon Nucleon scattering amplitudes of the Nijmingen group~\cite{Na79}. The only free parameter is the cut off mass of the dipole form factor which is used at each meson baryon vertex: the chosen value, $\Lambda_m= 1$ GeV, acts as an overall normalization factor and falls within the range of accepted values in this kind of game. 
Above 2 GeV, the variation with energy and the relative strength of each of the three channels are well accounted for: more particularly the predicted ratio of the cross sections of the $\Lambda$ and the $\Sigma^0$ production channel is about 3, in good accord with experiment. Also   the $\Sigma$ production channels are found to be dominated by $\pi$ meson exchange, while the $\Lambda$ production channel is found to get contribution from both $\pi$ and $K$ meson exchanges.

This simple model leads also to a excellent agreement with the spectra of the Kaons, emitted at different angles in the reaction $pp \rightarrow K^+ X$, accurately measured at SATURNE~\cite{Fra92}.
Above the Plane Wave contribution, the low energy $\Lambda P$ scattering reproduces the characteristic enhancement, near the end point of the spectrum, while  the coupling between the $\Lambda$ and $\Sigma$ channels is responsible for the narrow structure near the $\Sigma$  production threshold. The model reproduces also the angular distribution of these spectra.

\begin{figure}[h]%[b!] % fig1
\centerline{\epsfig{file=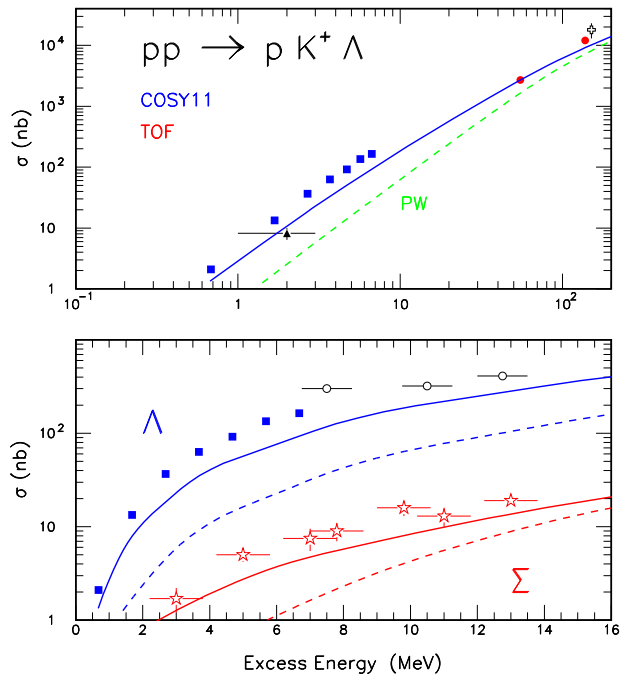,height=3.5in}}
\vspace{-1cm}
\caption{The total cross-sections recently determined at COSY. Dashed lines: Plane Wave. Full lines: FSI.}
\label{cosy}
\end{figure}

At COSY, precise determination of the cross sections of strangeness production channels are possible over a wider part of the phase space  than at SATURNE. Accurate data have been recorded, both for the $\Lambda$~\cite{Ba98} and $\Sigma^0$~\cite{Se99} production channels, very close to their respective threshold.  Fig.~\ref{cosy} summarizes these results. Without any adjustment, the simple model~\cite{La91} reproduces fairly well not only the shape and magnitude (over four decades), but also the ratio between the cross-sections of the $\Lambda$ and $\Sigma^0$ production channels. This ratio varies from about 27 near threshold to about 3 at high energy (above 2 GeV) ! 
Near threshold the model underestimates the data by about 50\%: This can be easily accommodated by adjusting the cut-off mass of the hadronic form factors. I did not play this game, since care has to be taken not to spoil the good agreement with the SATURNE data~\cite{Fra92}.

\begin{figure}[] % fig2
\vspace{-1cm}
\centerline{\epsfig{file=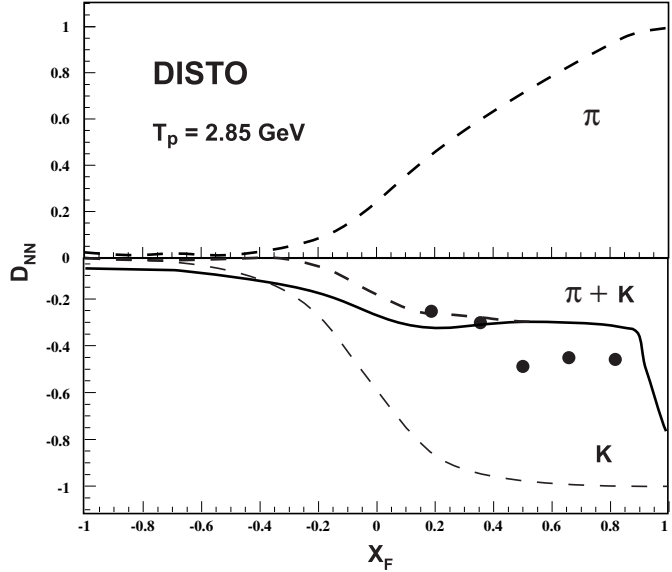,height=3.5in}}
\vspace{-1cm}
\caption{Spin transfer coefficient $D_{NN}$ as measured at Saturne. Dashed lines: Plane Wave. Full line: FSI.}
\label{disto}
\end{figure}

The balance between $\pi$ and $K$ exchange has been beautifully confirmed by the last experiment performed at SATURNE, DISTO~\cite{Ba99}. Among other things, the polarization transfer coefficient, $D_{NN}$, between the incident proton and the $\Lambda$ was determined  in the reaction $\vec{p}p \rightarrow p K^+ \vec{\Lambda}$. It is plotted in Fig.~\ref{disto}, against the Feynman variable $X_F= P_{\Lambda}/(P_{\Lambda})_{max}$. When $X_F=1$, {i.e} when the $\Lambda$ is emitted in the direction of the incident beam with the highest momentum, $\pi$ exchange leads to $D_{NN}= +1$, while $K$ exchange leads to  $D_{NN}= -1$. Without any adjustment, the simple model~\cite{La91} predicts $-0.3$, in close agreement with the data.

These findings show that the basic mechanisms are understood, and that the simplest model works. However, discrepancies of the order of 50\% still remain in the energy variation of the 
cross sections. This may be due to the neglect of a possible small contribution of $\rho$ and $K^*$ exchange. This may also reflect the effects of Unitarity constraints. The reason why the simple model~\cite{La91} works is that the exchange of spin zero mesons saturate the cross section. The corresponding Born Plane Wave amplitudes do not diverge when the energy increases: it is therefore easy to  simulate Unitarity constraints by using form factors at the various meson baryon vertices, and supplementing the Born term with Nucleon-Hyperon rescattering. This is not the case for the spin 1 (or higher) exchange amplitudes, which diverges with the energy: The use of form factors does not help to cure the problem, and this is the reason why Regge Poles have been introduced.
Indeed a Regge Pole description~\cite{So92} leads to a fair understanding of both the angular distribution and the $\Lambda$ polarization in the $pp \rightarrow p K+ \Lambda$ reaction at 400 GeV, again in terms of the elementary amplitudes of the  production of Hyperons by mesons.

The implementation of Regge Poles remains to be done at low energy. As I shall discuss now, this has been achieved in the Electromagnetic Sector.

Near  the Hyperon production thresholds, hadronic cross-sections are sensitive to Hyperon- Nucleon amplitudes. However a better understanding of the reaction mechanism has to be achieved.  The Electromagnetic Sector is in better shape in this respect.

\section{Electromagnetic Sector}

Following the successful description of pion photoproduction in the $\Delta$ and the second resonance region, Effective Lagrangian Models start from Born Terms in the tree approximation. However, when Kaon Nucleon coupling constants consistent with SU3 are used, they badly overestimate the experimental cross sections of Hyperon electroproduction channels. The reason is that, contrary to the pion production channels, the energy is higher, more coupled channels are open and unitarity constraints are more difficult to implement. It is possible to accommodate the models with the sparse available data, at the expenses of many additional contributions, which interfere destructively: baryonic resonance exchanges in the $s$ as well as in the $u$ channels, strange mesonic resonances in the $t$ channel, form factors at the various meson baryon vertices.

\begin{figure}[]%[b!] % fig3
\centerline{\epsfig{file=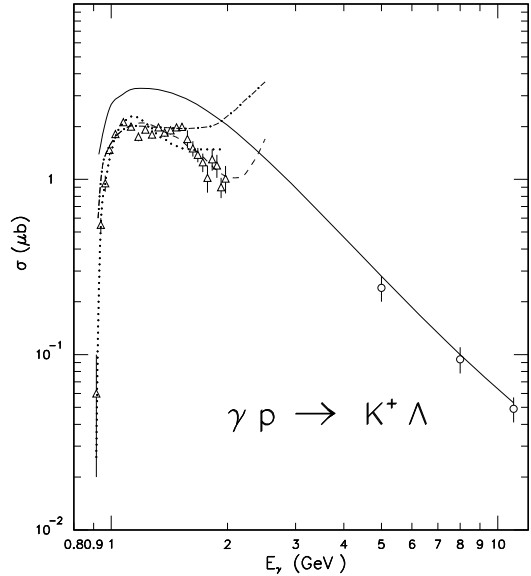,height=3.in}}
\vspace{-1cm}
\caption{The total cross-section of the $\gamma p\rightarrow K^+\Lambda$ reaction. Dot-dashed line: Lagrangian model, without hadronic form factors and off-shell effects~\protect\cite{SL96}. Dashed line: with off-shell effects~\protect\cite{Sa98}. Dotted line: with form factors~\protect\cite{Ha98}. Full line: Regge model~\protect\cite{La97}.}
\label{ELSA}
\end{figure}

This is illustrated in Fig.~\ref{ELSA} where typical results from  Effective Lagrangian Models~\cite{SL96,Sa98,Ha98} are compared to recent data form ELSA~\cite{Ta98} up to 2 GeV (CLAS preliminary data~\cite{Sch99} confirm them), and older data~\cite{Bo69} at higher energies.  While each reproduces the data in the  1--2 GeV range (they have been tailored for that) they diverge above 2 GeV. The use of form factors at the hadronic vertices helps, but does not cure the disease since it only postpones the divergence at higher energy. The reason is that, in order to fit the data (and since many degree of freedom are available), these models rely heavily on the exchange of particles (mesons in the $t$ channel, baryons in the $s$ and $u$ channels) with non zero spin. The corresponding Born amplitudes are known to diverge with energy. This is the reason why Regge Models have been introduced. Indeed, a recent application~\cite{La97} reproduces both the energy variation and the magnitude of the cross sections for strangeness production off the nucleon above 2 GeV. Below,  it overestimates the ELSA data, but here it is meant to average over the contributions of the various possible resonances (duality). 

They provides us with an economical and elegant way to describe meson photoproduction at high energies. Analyticity of the scattering amplitude is built in. The exchange of  families (Regge Trajectories) of mesons, in the $t$ channel, and baryons, in the $u$ channel,  ensures unitarity, and describes the off-shell behavior of the amplitudes far from the poles.  The striking feature is that they exhibit the right energy and momentum dependencies.

Taking advantage of a better knowledge of coupling constants, and of a larger data set, we have revived a Regge Model which reproduces, without free parameters, cross sections and polarization observables, available at high energies, in pion and kaon photoproduction channels~\cite{La97}  as well as electroproduction channels~\cite{Va98,Gui99}. It relies on the full spin structure of the elementary vertices, and is based on the exchange of $K$ and $K^*$ trajectories. I refer the reader to these works for a comprehensive review. Let me only flash two recent results: they concern spin observables and constitute  a more stringent check of the model.

\begin{figure}[]%[b!] % fig4
\centerline{\epsfig{file=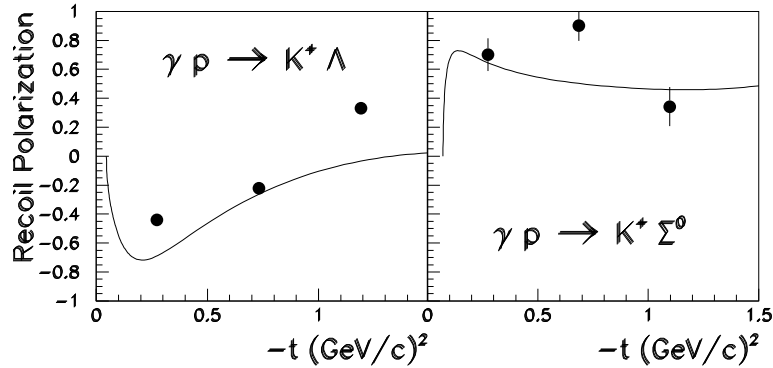,width=4.in}}
\vspace{-1cm}
\caption{Recoil Polarization of the $\Lambda$ or the $\Sigma$
emitted in the reaction $\gamma p\rightarrow K^+ Y$.}
\label{recoil}
\end{figure}

Fig.~\ref{recoil} shows the recoil polarization of the hyperon recoiling in the reaction $p(\gamma, K^+) Y$, recently determined at ELSA~\cite{Ta98}. At low $t$ the Regge model~\cite{Gui99} reproduces fairly well the data, and predicts the right sign in both the $\Lambda$ and $\Sigma$ channels. This spin observable is particularly sensitive to the modeling of the reaction, since it depends on the imaginary part of the amplitude, and is usually difficult to reproduce. It comes out naturally in the Regge model.

\begin{figure}[]%[b!] % fig5
\centerline{\epsfig{file=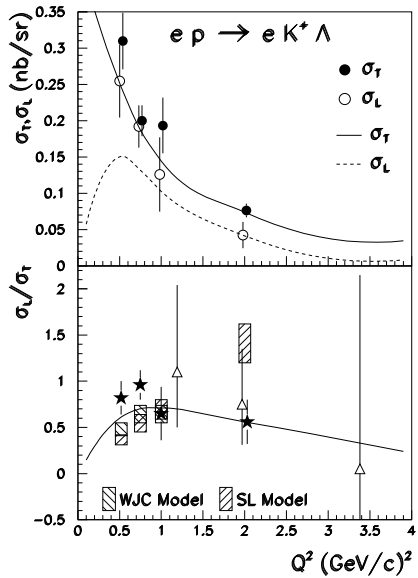,height=3.in}}
\vspace{-1cm}
\caption{The Transverse and Longitudinal cross-sections, and their ratio, as recently measured at JLab in the reaction $p(e,e'K^+)\Lambda$. The curves are the predictions of the Regge model. The hashed areas are the predictions of Effective Lagrangian Models.}
\label{cebaf}
\end{figure}

Fig.~\ref{cebaf} shows the Transverse and Longitudinal cross sections of the $p(e,e'K^+)\Lambda$ reaction, which have been recently determined at JLab~\cite{Ni98} through a Rosenbluth separation. While Lagrangian Models~\cite{SL96,Wi92} predict a linear rise of their ratio with the four momentum transfer $Q^2$ of the virtual photon, the Regge model~\cite{Gui99} correctly reproduces the saturation and the decrease of the ratio at higher virtuality of the photon.

To summarize, at low energies (let say below 2 GeV) where many resonances contribute, only a comprehensive  experimental determination of the reaction amplitudes (partial wave analysis) will allow to decipher the various contributions  and constrain Lagrangian models. On the theoretical side, a serious effort has to be done in order to implement Unitarity  constraints in  highly coupled channels.

At high energies (above 2 GeV), Regge Model provides us with a successful, simple  and economic way to describe data at forward angles ($t\leq 1$ GeV) and backward angles ($u\leq 1$ GeV). In between (around $90^{\circ}$) there is room for hard scatterings, which can be described by saturating Regge trajectories~\cite{La97}.

\section{Few Body Systems}

\begin{figure}[h]%[b!] % fig6
\centerline{\epsfig{file=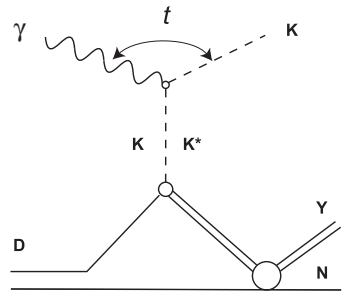,height=1.5in}}
\vspace{-1cm}
\caption{The  rescattering graph. }
\label{graph}
\end{figure}

Such a Regge model of the elementary amplitude can be easily implemented in Few Body systems, following the method~\cite{La81} which has been successfully used in the pion photoproduction sector. Among others, the determination of the Hyperon-Nucleon scattering amplitudes appears to be particularly appealing at low as well as high energy. The typical diagram is depicted in Fig.~\ref{graph}: the elementary Kaon production amplitude and the deuteron wave function are well under control. It provides the most direct way to access Hyperon-Nucleon scattering, in the absence of  Hyperon beams. Furthermore the determination of the polarization of the emitted Hyperon provides us with a strong additional constraint. 

\begin{figure}[]%[b!] % fig7
\centerline{\epsfig{file=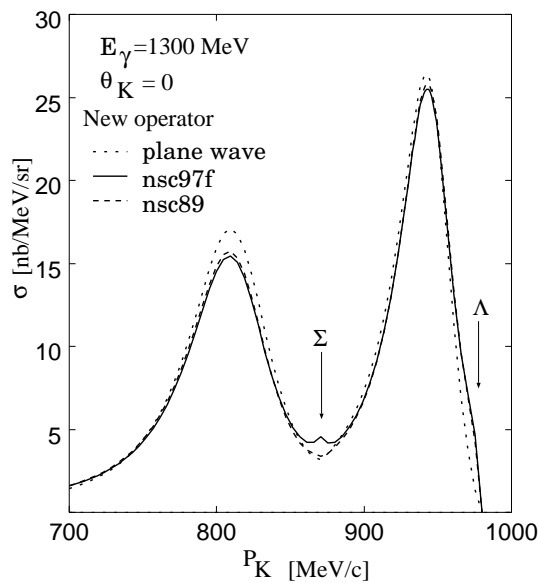,height=3.in}}
\vspace{-1cm}
\caption{Theoretical cross-section\protect\cite{Be99} of the  $D(\gamma, K^+)  Y n$ reaction. }
\label{bennhold}
\end{figure}

\begin{figure}[]%[b!] % fig8
\centerline{\epsfig{file=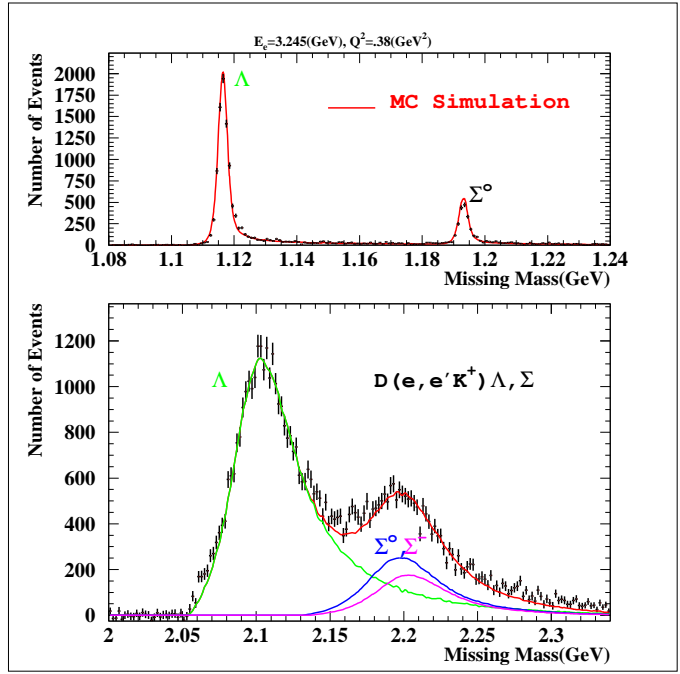,height=3.5in}}
\vspace{-0.5cm}
\caption{The  yield of the  $D(e,e' K^+) Y n$ reaction at CEBAF~\protect\cite{Re99}.  Top panel: proton target.}
\label{d_cebaf}
\end{figure}

Low energy YN rescattering leads to characteristic enhancements in the spectrum of the Kaons, emitted in the reaction $D(\gamma,K^+) Y n$, near the $\Lambda$ and $\Sigma$ production thresholds. This is clearly apparent in Fig.~\ref{bennhold}, which shows the latest prediction for a kinematics accessible at CEBAF~\cite{Be99}.  This picture is confirmed by a recent measurement shown in Fig.~\ref{d_cebaf}. However, the effects are much less dramatic than in proton scattering~\cite{La91}. The reason is that the dominant mechanism is the quasi-free production of Kaons on a nucleon almost at rest in Deuterium. This mechanism is responsible for the two broad peaks in Figs.~\ref{bennhold} and~\ref{d_cebaf} ($\Lambda$ and $\Sigma$ production), of which the width is due to the Fermi motion of the target nucleon inside the Deuteron. $\Lambda n$ and $\Sigma^{\circ}n$ rescatterings appear on the tails of the quasi free peaks. The way to overcome this difficulty consists in performing an exclusive experiment where one select high momenta of the recoiling neutron, in order to suppress quasi-free mechanisms. This has become possible thanks to the high duty factor, the high luminosity of CEBAF and the large acceptance of CLAS. 

Such exclusive measurements, at higher energies, allow also to reach an interesting kinematical regime where both the nucleon and the Hyperon propagate on-shell in the intermediate state. The corresponding triangular singularity induces a narrow peak on top of the quasi-free contribution, in well defined part of  the phase space.  The rescattering amplitude depends only on on-shell amplitudes and on the low momentum components of the Deuteron wave function. I refer the reader to Refs.~\cite{La98,La99} for a detailed discussion of these issues.

The pion production sector provides us with a solid testing ground of this conjecture. In the kaon production sector, this particular kinematics gives access to the on-shell Hyperon-Nucleon scattering amplitude. Two regime are interesting. At low momentum transfer $t$, between the incoming photon and the outgoing Kaon, the propagating Hyperon is not too much affected and one access the ``normal'' on-shell Hyperon-Nucleon scattering amplitude. At high momentum transfer, the constituents of the Hyperon are forced to stay in the small interaction volume and a ``small'' Hyperon propagates experiencing Color Transparency. To what extent the presence of a strange quark will affect Color Transparency is a fascinating open question.

\section{Hidden Strangeness Production}

\begin{figure}[]%[b!] % fig9
\centerline{\epsfig{file=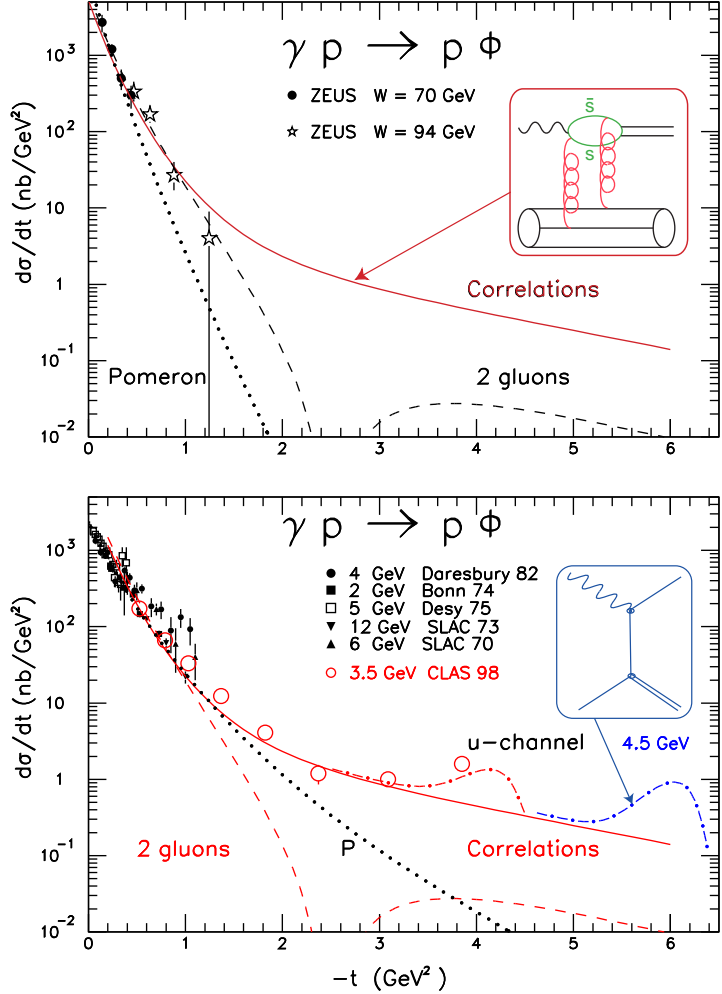,height=5.5in}}
\vspace{-1.cm}
\caption{The differential cross section of the $p(\gamma, \phi)p$ at HERA (top panel) and JLab (bottom panel).}
\label{phi}
\end{figure}

Hidden strangeness photoproduction allows to prepare a pair of strange antistrange $s\bar s$ quarks, such as the $\phi$ meson, and study its interaction with hadronic matter. At low momentum transfer $t$ (small angle), its diffractive scattering is mediated by the exchange of the Pomeron. At high momentum transfer (large angle), the impact parameter is small and comparable to the gluon correlation length (the distance over which a gluon propagates before hadronizing): the Pomeron is resolved into its simplest component, two gluons which may couple to each of the quarks in the emitted vector meson or in the proton target. This is illustrated in Fig.~\ref{phi}, which shows data recently recorded at DESY~\cite{Br00} and JLab~\cite{An00}. At low $t$, the data confirm the shrinkage of the forward diffraction peak and the slow rise of the cross section with the energy, as expected from the exchange of the Pomeron Regge trajectory. The two gluon exchange contribution matches the Pomeron exchange contribution around $-t\sim 1$GeV$^2$ and reproduces the data at higher $t$. In the JLab energy range, $u$-channel nucleon exchange ``pollutes'' the highest $t$ bin: here the $\phi NN$ coupling constant $g_{\phi NN}= 3$  is the same as in the analysis of the nucleon electromagnetic form factors~\cite{Ja89}. More details may be found in ref.~\cite{La00}.

Such experiments open a window on the study of Van der Waals force betwen hadrons.

\section{Conclusion}

The production and the propagation of the strange quark provides us with an ``intruder'' which probes the inner structure of hadronic matter. Such a opportunity has not been fully exploited in the past, due to the low intensity of the available beams. This is changing with  the advent of new facilities (COSY, JLab, HERMES, COMPASS), where strangeness production reactions are undergoing a renewed interest. 

On the theoretical side, the main players in the game are identified. At intermediate energies (COSY, JLab), a Regge description is best suited to the description of reaction mechanisms. It has already been implemented in the electromagnetic sector; this remains to be done in the hadronic sector. At high energies, the detection of Kaons or $\Lambda$ tags the interaction of the probe with a strange quark. However a good understanding of the hadronization chain must be achieved.

I am pretty confident that many new, experimental as well theoretical, results will be reported at the next Conference. Surprises may even happen in the mean time!


\begin{thebibliography}{99}

\bibitem{Be91} R. Bertini, {\it Europhysics News } {\bf 22}, 115 (1991).

\bibitem{Fla79} V. Flamino {\it et al.}, {\it Compilation of Cross Sections} {\bf CERN-HERA 79-03} (1979).

\bibitem{La93} J.-M. Laget, {\it Journal of the Korean Physical Society} {\bf 26}, S244 (1993).

\bibitem{La97} M. Guidal, J. M. Laget and M. Vanderhaeghen, {\it Nucl. Phys.} {\bf A627}, 645 (1997).

\bibitem{La91} J.-M. Laget, {\it Phys. Lett.} {\bf B259}, 24 (1991).

\bibitem{Na79} M. M. Nagels, T. A. Rijken and J. J. de Swart, {\it Phys. Rev.} {\bf D15}, 2547 (1977); \\{\bf D20}, 1633 (1979). 

\bibitem{Fra92} R. Siebert {\it et al., Nucl. Phys.} {\bf A567}, 819 (1994).

\bibitem{Bi98} R. Bilger {\it et al., Phys. Lett.} {\bf B420}, 217 (1998).

\bibitem{Ba98} J. T. Balevsky {\it et al., Phys. Lett.} {\bf B420}, 211 (1998).

\bibitem{Se99} S. Severin {\it et al., Phys. Rev. Rev.} {\bf 83}, 682 (1999).

\bibitem{Ba99} F. Ballestra {\it et al., Phys. Rev. Lett.} {\bf 83},  1534 (1999).

\bibitem{So92} J. Soffer and T\"{o}rnquist, {\it Phys. Rev. Lett.} {\bf 68}, 907 (1992).

\bibitem{SL96} J. C. David {\it et al., Phys. Rev.} {\bf C53}, 2613 (1996).

\bibitem{Sa98} T. Mizutani {\it et al., Phys. Rev.} {\bf C58} 75 (1998).

\bibitem{Ha98} H. Haberzettl {\it et al., Phys. Rev.} {\bf C58}, R40 (1998).

\bibitem{Ta98} M. Q. Tran {\it et al., Phys. Lett.} {\bf B445}, 20 (1998).

\bibitem{Bo69} A. M. Boyarski {\it et al., Phys. Rev. Lett.} {\bf 22} 1131 (1969). 

\bibitem{Sch99} R. Schumacher {\it et al.}, Communication to PANIC99. 

\bibitem{Va98} M. Vanderhaeghen, M. Guidal and J.-M. Laget, {\it Phys. Rev.} {\bf C57}, 1454 (1998).

\bibitem{Gui99} M. Guidal, J.-M. Laget and M. Vanderhaeghen, {\it Phys. Rev.} {\bf C61}, 025204 (2000).

\bibitem{Ni98} G. Niculescu {\it et al., Phys. Rev. Lett.} {\bf 81}, 1805 (1998).

\bibitem{Wi92} R. A. Williams {\it et al., Phys. Rev.} {\bf C46}, 1617 (1992).

\bibitem{La81} J.-M. Laget, {\it Phys. Rep.} {\bf 69}, 1 (1969).

\bibitem{Be99} H. Yamamura {\it et al., Phys. Rev.} {\bf C61}, 014001 (1999). 
 

\bibitem{Re99} D. Abbott {\it et al.,  Nucl. Phys.} {\bf A639}, 197c (1999), and private communication.

\bibitem{La98} J.-M. Laget, {\it Proceedings of the Worskshop ``Physics and Instrumentation with 
6--12 GeV Beams''}. Eds. S. Dytman {\it et al.}, JLab User Production, 1998, pp. 57.

\bibitem{La99} J.-M. Laget, {\it Proceedings of the Workshop ``Exclusive and Semi-Exclusive Processes at High Momentum Transfer''} (JLab, USA, May 1999). Eds. C. Carlson and A. Radyushkin, World Scientific, 1999, p. 13.

\bibitem{Br00} J. Breitweg {\it et al., Eur. Phys. J.} {\bf C14}, 213 (2000).

\bibitem{An00} E. Anciant {\it et al., Phys. Rev. Lett.} {\bf 85}, 4682 (2000).

\bibitem{Ja89} R. L. Jaffe, {\it Phys. Lett.} {\bf B229}, 275 (1989).

\bibitem{La00} J.-M. Laget, {\it Phys. Lett.} {\bf B489}, 313 (2000).

\end{thebibliography}
\end{document}